\documentclass{aa}  
\usepackage{graphicx}
\usepackage{txfonts}
\usepackage{amssymb}
\usepackage{hyperref}
\usepackage{color}
\newcommand{\eg}{\emph{e.g.}}
\newcommand{\ie}{\emph{i.e.}}

\begin{document}

\title{Changes in granulation scales over the solar cycle seen with SDO/HMI and \emph{Hinode}/SOT}
\author{J. Ballot\inst{1}
\and
        T. Roudier\inst{1}
\and
       J.M.~Malherbe\inst{2}
\and
       Z.~Frank\inst{3}
}
\offprints{J. Ballot,\\
\email{jerome.ballot@irap.omp.eu}}

\institute
    {
      Institut de Recherche en Astrophysique et Plan\'etologie (IRAP), Universit\'e de Toulouse, CNRS, UPS,
      CNES, 14 avenue Edouard Belin, 31400 Toulouse, France
\and
Observatoire de Paris, LESIA, 5 place Janssen, 92195 Meudon, France, PSL Research University, CNRS, Sorbonne Universit\'es,
UPMC Univ. Paris 06, Univ. Paris Diderot, Sorbonne Paris Cit\'e 
\and
Lockheed Martin Solar and Astrophysics Laboratory, Palo Alto, 3251 Hanover Street, CA 94303, USA
}

\date{Received 15 September 2020; Accepted 25 May 2021}
%\titlerunning{}
%\authorrunning{}

%Context
%Aims
%Method
%Results
%Conclusions

\abstract
{The Sun is the only star where the superficial turbulent convection can be observed at very high spatial resolution.
  The Solar Dynamics Observatory (SDO) has continuously observed the full Sun from space with multi-wavelength filters since July 2010. In particular, the Helioseismic and Magnetic Imager (HMI) instrument
  takes high-cadence frames (45 seconds) of continuum intensity in which solar granulation is visible.}
{We aimed to follow the evolution of the solar granules over an activity cycle and look for changes in their spatial properties.}
{We investigated the density of granules and their mean area derived directly from the segmentation of deconvolved images from SDO/HMI. To perform the segmentation, we define granules as convex elements of images.}
{We measured an approximately 2\% variation in the density and the mean area of granules over the cycle, the density of granules being greater at solar maximum with a smaller granule mean area. The maximum density appears to be delayed by about one year compared to classical activity indicators, such as the sunspot number. We complemented this study with high-spatial-resolution observations obtained with \emph{Hinode}/SOTBFI (Solar Optical Telescope Broadband Filter Imager), which are consistent with our results.}
{The observed variations in solar granulation at the disc centre reveal a direct insight into the change in the physical properties that occur in the upper convective zone during a solar cycle. These variations can be due to interactions between convection and magnetic fields, either at the global scale or, locally, at the granulation scale.}


\maketitle

\section{Introduction}

Contrary to other stars, we are close enough to the Sun to be able to observe various physical properties of its surface with high spatial resolutions (up to 180 km) and high temporal cadences (5 to 60\:s). Convective elements, known as 
solar granules, are visible from ground-based and space-based observatories. Satellites provide images without the terrestrial 
atmospheric distortions (seeing) and allow precise quantitative measurements of granule physical properties.
Although the main characteristics of that granulation are widely described in the literature  \citep[\eg\ see ][]{HBS2002,NSA2009,LSVD2014,Fischer2017,FPGR2017}, the studies of its
variations over activity cycles are limited \citep[see references in][]{RR1998,Muller2018}. This is mostly due to the difficulty in acquiring homogeneous sets of granulation images without atmospheric distortions over a whole 11-year solar cycle.
Another method for tracking variations in granulation properties consists in analysing the power spectrum of solar radial-velocity or luminosity fluctuations. Using GOLF data, \citet{Lefebvre08} looked for variations in solar granular timescales with the activity cycle. However, they did not find any significant variations correlated to the magnetic activity because they were dominated by instrumental and orbital effects.

Since July 2010, the Helioseismic and Magnetic Imager (HMI) instrument aboard the Solar Dynamics Observatory (SDO) spacecraft has provided images of solar granulation free of atmospheric perturbations with a moderate spatial resolution and a temporal cadence of 45 sec. These data provide an opportunity to determine the variation amplitude of granulation parameters at the solar surface over several years.
In addition, we had access  to high-resolution images of the solar granulation obtained with the 50 cm Solar Optical Telescope (SOT) aboard the \emph{Hinode} satellite between November 2006 and February 2016.

Complex interactions between solar turbulent convection and magnetic fields generate the quasi-periodic solar activity cycle. The appearance of strong magnetic structures (spots, pores, etc.) at the photosphere level is known to locally modify the convection processes. 
\citet{Faurobert18} have shown that activity-induced variations in physical quiet-Sun structures, such as the temperature gradient, might affect the irradiance.
In this study we look for global changes in granulation that would also affect the quiet Sun.
Variations in solar granulation properties would provide a direct insight into the
change in physical properties occurring in the upper convective zone. Quantifying these variations along a cycle would provide new observational constraints on interactions between convection and the magnetic field.

It has long been established that variations in solar granulation represent only a few tens of parts per million of solar luminosity variations \citep{HUDWOOD1983}.
Nevertheless, we can observe the evolution of the dynamical properties of the granulation over a solar cycle. Thanks to \emph{Hinode} observations, variations in intensity contrast and in granulation scales have been shown to be smaller than 3\% \citep{Muller2018}. 

\begin{figure*}[!htbp]
 \centerline{%
\includegraphics[width=0.33\hsize,clip=]{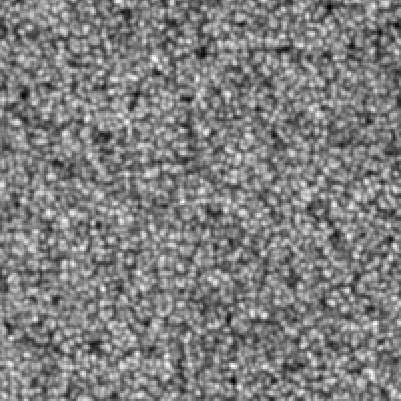}%
\includegraphics[width=0.33\hsize,clip=]{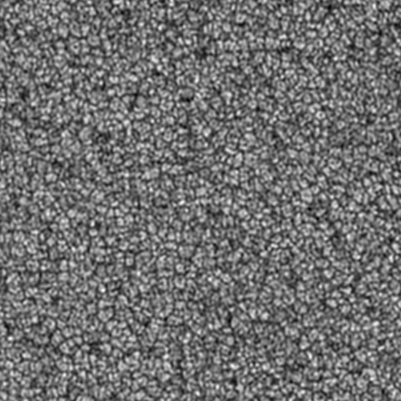}%
\includegraphics[width=0.33\hsize,clip=]{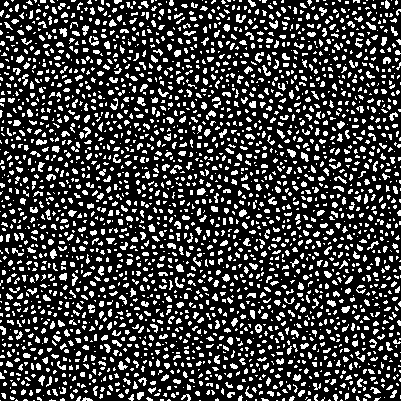}}
 \caption{Progression of the central part of a processed image. \emph{(Left)} Original image of solar granulation observed with HMI around the disc centre with a pixel size of 0\farcs5\ for a field of view of $100\arcsec \times 100\arcsec$. \emph{(Middle)} Same image after deconvolution, with a pixel size of 0\farcs25. \emph{(Right)} Same image, segmented at the granulation level (each white element corresponds to a granule).
 \label{fig:images}}
 \end{figure*}

We aim to go further with the present study, taking advantage of SDO/HMI observations and using a new approach. We investigate the density of granules and their mean area, deriving them directly from segmented images of the photosphere provided by SDO/HMI. In Sect. \ref{sec:red} we briefly describe the data selection and the reduction pipeline we used to get the segmented images. In Sect. \ref{sec:evol} we present the temporal evolution of granule geometrical parameters and discuss the corrections we made. We compare our results to complementary observations provided by \emph{Hinode}/SOT in Sect.~\ref{sec:hinode}. In Sect.~\ref{sec:mag} we determine the contribution of the magnetic network, before concluding in Sect.~\ref{sec:conc}.

\section{SDO/HMI data selection and reduction}\label{sec:red}

The HMI instrument \citep[][]{Scherrer2012,Schou2012} aboard SDO provides uninterrupted observations over
the entire solar disc. This provides a unique opportunity to extract solar granulation characteristics over a long time period with uniform observation sets.

Since data reduction is very time consuming, it was not possible to reduce all available data; as such, we had to carefully pick our image set. We selected one day per month over nine years, starting on 1 July 2010, with a time step as regular as possible from available data in the Joint Science Operations Center (JSOC) data base. Such a monthly selection allows us to check the orbital effects of the SDO satellite, which introduce annual variations.
For each selected day, we picked hourly images. We ensured, via visual inspection, that all images have good quality. We thus ended up with a set of 2711 SDO/HMI white-light images.

The original pixel sizes of all the images are close to 0\farcs5; the exact values were used to reduce and analyse the data. White-light $4096\times4096$ images were deconvolved from the HMI
transfer function \citep{Couv2016} and re-binned by a factor of two, resulting in images with a size of $8192\times 8192$. This operation allowed us to increase the number 
of pixels for each granule and aided in the segmentation processing. Such a technique has already been successfully applied in \citet{Roudier2020}.
To limit projection effects, the field of view was restricted to $1000\times1000$ pixels around the disc centre. This corresponds to about $250\arcsec\times250\arcsec$,
with a slight dependence on the distance to the Sun during the orbit.
The conversion from arcseconds to kilometres takes into account the exact distance between the SDO satellite and the Sun.

Different techniques for extracting granules from each image have been proposed over the years and were recently discussed by \citet{Roudier2020}. For this current work, we chose to extract each granule by using convexity detection, as done in \citet{Roud2012}. With this method, granules are identified as convex elements of the image. Contrary to other approaches, such as watershed methods or the method developed by \citet{BW01}, the technique based on convexity is, by construction, independent of the image contrast and does not require free parameters that would have to be manually tuned for different images. We thus applied a very homogeneous treatment to all the images over nine years. A major flaw of the convexity method is that it underestimates the granule size. This would have been an issue if we wanted to measure absolute quantities, but it did not affect our study since we looked for relative variations.

Due to our selection method (one frame per hour), five-minute oscillations are still present in the data and may affect our analysis.
We thus performed tests on a time sequence with a 45 s time step to be able to filter out the five-minute oscillations, as done for example in \citet{Roudier2019}. We thus verified that our segmentation technique is only weakly sensitive to these oscillations.

Once the images were segmented, the final steps consisted in labelling and counting the granules.
To reduce the impact of noise,
we discarded convex elements with an area of one pixel, that is, we only counted as granules convex elements with an area greater than or equal to two pixels.
Figure~\ref{fig:images} shows the progression of the central part of a processed image, from the raw HMI data to the final image -- segmented at the granulation level -- via the deconvolved image. 

 \section{Solar granule density and area over a solar cycle}\label{sec:evol}

The deconvolved and segmented images, at the solar disc centre, allowed us to get the number of granules and their areas in the observed field of $1000\times1000~$ pixels and to follow them over time. We detail in this section how we extracted the granule density and granule area and how we corrected them for annual effects.
 
\subsection{Temporal evolution of the granule density}\label{ssec:evol_density}

Counting the number of granules in each frame was done by labelling the segmented images. Of the 2711 values, four are clear outliers (> 10-$\sigma$); they were removed from the sample and are not considered in the rest of the article. Thus, our final set was reduced to 2707 points.
Due to the combination of the geostationary and terrestrial orbit of the 
SDO satellite, the real field of view in square megametres (Mm$^2$) on the Sun's surface varies by a few fractions of a percent over a year.
The exact area in Mm$^2$ of the observed field for each frame was determined  by using the distance of SDO to the Sun given in the FITS file headers. The density of granules per Mm$^2$ was then computed as the counted number of granules divided by the exact observed area in Mm$^2$.
 
We needed to properly treat the edges of images since granules on image boundaries are cut, which impacts the measured densities. To correct for this effect, we defined a border zone as a six-pixel-wide band running along the boundaries of images: Granules that were fully inside this zone were discarded, whereas granules that straddled the frontier of this zone were counted with a weight representing the fraction of the granule inside the image. For example, if 60\% of the area of a granule is inside the boundary zone and 40\% inside the image, this granule is count as 0.4 when computing the density.

 Neglecting to correct for edge effects induces a systematic overestimation of the density of 0.2\%. Nevertheless, this bias is very stable from one image to another (image-to-image fluctuations are an order of magnitude smaller), and thus we are very confident that our correction does not introduce spurious variations in this study.

 \begin{figure}[!htbp]
 \includegraphics[width=\hsize]{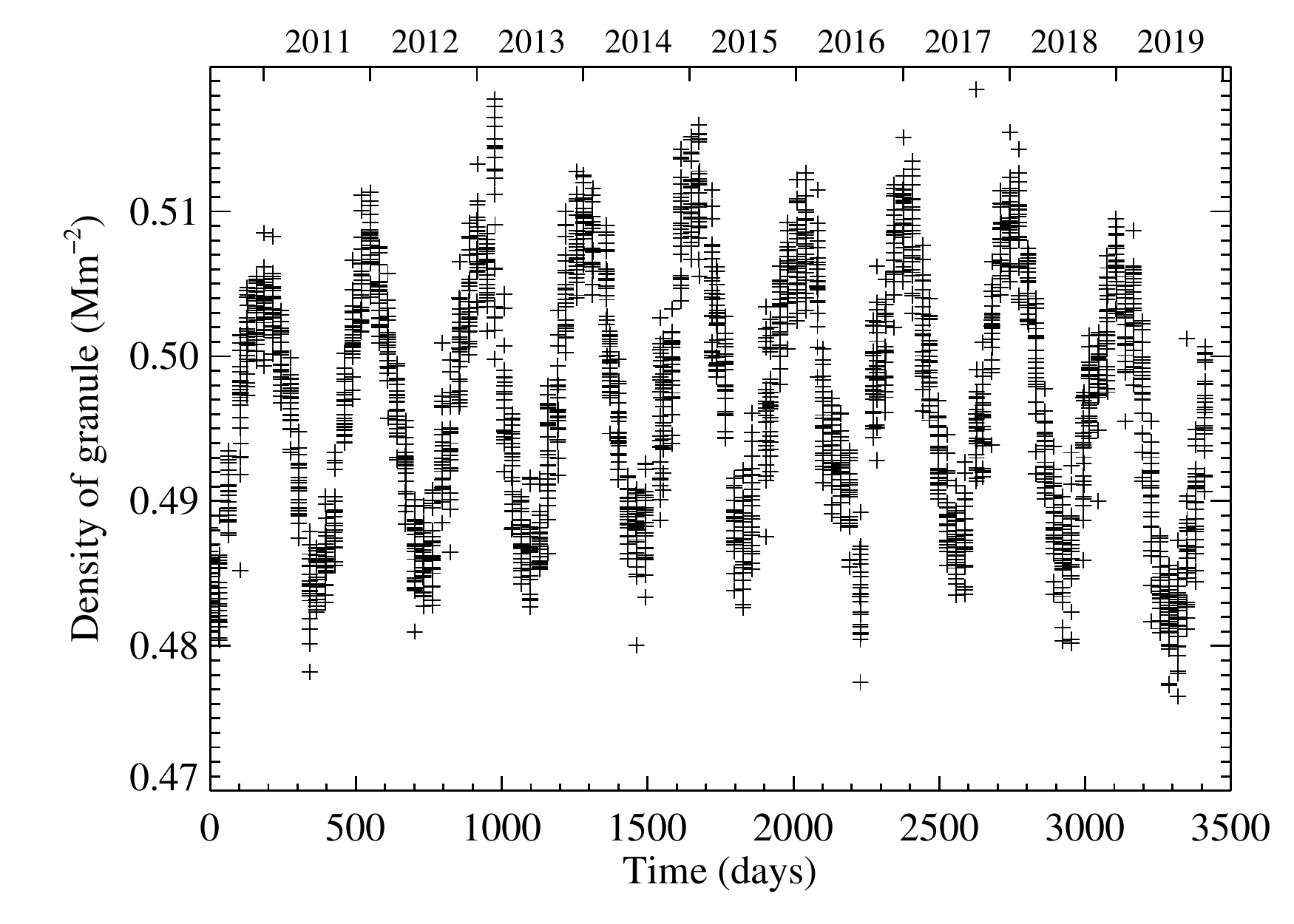}
\caption{Temporal evolution of the observed density of granules per Mm$^{2}$ measured at the disc centre with SDO. Each cross corresponds to the measurement obtained from an image. The time axis starts on 1 July 2010. The upper x-axis indicates observation years.}
\label{fig:evol_density}
\end{figure}

 Figure~\ref{fig:evol_density} shows the temporal evolution of the density of granules (in Mm$^{-2}$) during the cycle.
  There is still an annual variation in the amplitude of around 5\%, which is clearly visible in the density evolution despite the field-of-view size for each image being taken into account. Nevertheless, a global change over the cycle is visible: We notice an increase in the first half and a decrease in the second half.
  
We needed to understand the origin of this annual oscillation, which is undoubtedly due to the analysis method or the instrument itself. Thus, a detailed inspection of the data reduction procedure was required to determine which parameters can generate such a fluctuation. Combinations of the orbit velocity, CCD temperatures, and window temperature affect the measured filtergram intensities (P. Scherrer, 
private communication). Knowing this, we checked three parameters that could be involved in the observed residual  annual variations:
 (i) the image contrast, (ii) the focus, and (iii) the velocity of SDO/HMI relative to the Sun.

The first tested parameter was the contrast. Over their life cycle, the detectors of HMI show a decrease in sensitivity \citep[Fig.~6 in][] {Hoek2018}, which can also affect the contrast of the solar granulation. To check this, we normalized the contrast with the same mean for two days of the sequence and applied our reduction pipeline. As explained in Sect.~\ref{sec:red}, our segmentation method, which is based on convexity, is mathematically insensitive to the contrast. As expected, the numbers of granules detected with and without the contrast normalization are very similar. However, we must keep in mind that we only tested a change in contrast without loss in image quality. Blurred images (caused by \eg\ de-focus effects) are lacking in both contrast and sharpness. 

We then checked the density of granules relative to the image focus. During its journey, the SDO satellite receives direct solar flux as well as solar flux reflected by the Earth. The evolution of that total flux modifies the temperature of several elements 
 (CCDs, windows, etc.)  in the satellite, which in turn slightly changes the focus. 
 The uncontrolled window temperature is particularly known to affect the focus \citep{Hoek2018}.
 The quality of the focus is one of the parameters that could possibly influence our granule detection. To test that hypothesis, we mimicked de-focus effects by artificially degrading
 the modulation transfer function \citep{Couv2016}
of an image. The degraded image lost 1.85\% in contrast; this only slightly affected the number of detected granules, which was reduced by 0.45\%. It is difficult to push the blurring much further while still maintaining an acceptable image. Such a contrast variation (1.8\%) is representative of variations observed in real data over a day. Over a year, the daily mean contrast varies by 2$\sim$3\%. However, we must keep in mind that other parameters (\eg\  spatial resolution in megametres) also affect the contrast, and hence the variation in contrast due to de-focussing is necessary smaller than the total observed variation. Nevertheless, we found that when one increases the de-focussing to get 5\% fewer detected granules, the image contrast drops by at least 10\%, far beyond what is observed. Thus, changes in focus do not go far in explaining the observed annual variations; these changes may, however, contribute to the dispersion measured over a day.

\begin{figure}[!htbp]
\includegraphics[width=\hsize]{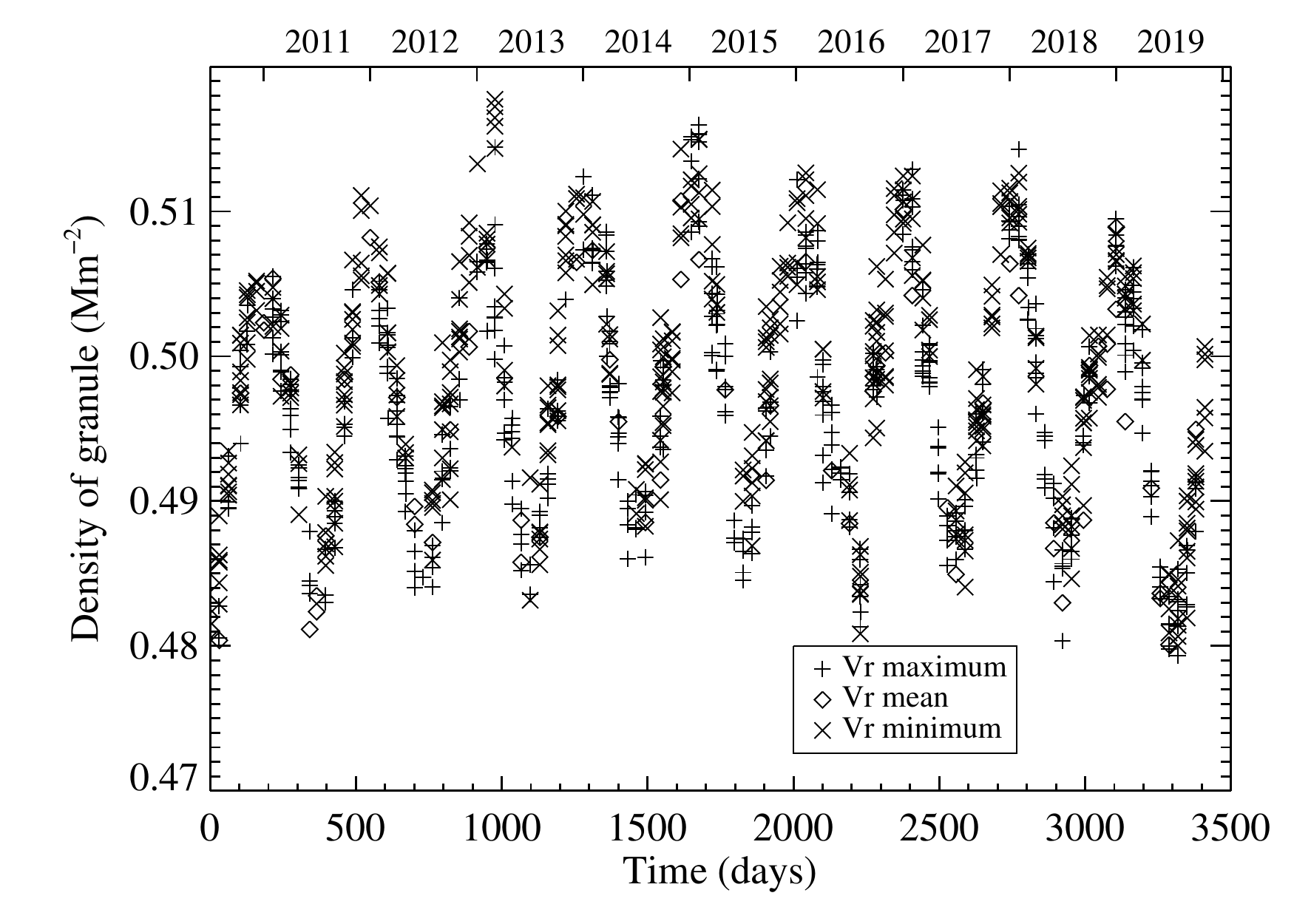}
\caption{Similar to Fig.~\ref{fig:evol_density} but the data have been split into three categories, corresponding to three different ranges of radial velocity, $V_r$, of the SDO satellite (maximum, mean, or minimum; see the main text for details).
\label{fig:density_vr}}
\end{figure}

Finally, intensity measurements could also be affected by the high velocity of SDO relative to the Sun, $V_r$. To check the impact of $V_r$, we selected three data subsets corresponding to three different ranges in velocity:
 maximum ($V_r > 2$\:km\,s$^{-1}$), mean ($-0.1<V_r<0.1$\:km\,s$^{-1}$), and minimum ($V_r < -2$\:km\,s$^{-1}$).
Figure~\ref{fig:density_vr} shows the density of granules for these three samples. The $V_r$ component does not appear to be involved in the observed annual variation since the measurements show a distribution over the whole curve, regardless of the amplitude of $V_r$.
 
During its journey, SDO does not observe the Sun from the same direction: The $B_0$ angle varies with a one-year period. This means that the field of view is centred on solar latitude -7.26\degr\ in early September and 7.26\degr\ in early March. We cannot imagine a physical reason linking $B_0$ to the measured granule number -- except if there were a never-before-seen strong dependence on the latitude. Moreover, the $B_0$ variation is not in phase -- nor in antiphase -- with the measured density. Indeed, every 6 months (early June and early December) the field of view is centred on the solar equator ($B_0=0\degr$), but the mean density is $\sim 0.49$ in June and $\sim 0.51$ in December. Thus, $B_0$ is not a relevant parameter for explaining the annual oscillation.

 \subsection{Impact of the pixel area}\label{ssec:pix}

Since the parameters described above did not allow us to understand the observed annual variation, we continued our inspection of the parameter space.
 A detailed inspection of the granule binarization revealed that the segmentation process, by convexity detection, is sensitive to the
 pixel size in megametres. This sensitivity is due to the fact that the smallest granules have a size close to that of the pixel, even after deconvolution and re-binning at 0\farcs25. Thus, with larger pixels, the segmentation method misses the smallest granules, which reduces the measured density -- and increases the measured mean area (see  Sect.~\ref{ssec:area}). We are close to the detection limit, but, as shown in Fig.~\ref{fig:images}, granules are clearly visible and segmented.
 
\begin{table}[!htbp]
\begin{center}
\begin{tabular}{lc}
  \hline
  \hline
  & Perih./Aph. \\
  \hline
   Mean granule area & $-3.3\%$\\
   Filling factor & $+0.6\%$\\
   Granule density & $+4.0\%$\\
  \hline
\end{tabular}
\caption{Variations in the determination of the mean granule area, the granule filling factor, and the granule density between two artificial observations of the same granulation simulation: one using a pixel area corresponding to SDO perihelion, the other to SDO aphelion.}
\label{tab1}
\end{center}
\end{table}
 
To quantify the effect of the pixel size on our segmentation, we used the solar granulation
 simulation performed by \citet{SN00} to generate artificial observations of the photosphere with two different pixel sizes, one corresponding to the SDO perihelion and the other to the aphelion. We then compared the mean granule area, the granule filling factor, and the granule density. Table~\ref{tab1} summarizes the comparison of these three quantities between these extreme positions. We observe a clear impact of the pixel size on the granule density and mean area due to the segmentation processing. The amplitude of the granule density variation reported in Table~\ref{tab1} is the same as the one observed in the real data.
We conclude that the changes in pixel size with the distance of SDO to the Sun generate the artificial oscillation visible in Fig.~\ref{fig:evol_density}.

\subsection{Corrected granule density}\label{ssec:density}

Due to the sensitivity of our measurement on the pixel size, we cannot infer any conclusions about variations shorter than one year. However, we can directly compare the granule density obtained from images with the same pixel size. Thus, we can recover relevant variations over a decade since, year after year, at the same epoch the pixel size is the same. As a consequence, the global variation of about 1-2\% visible in Fig.~\ref{fig:evol_density} is not explained by pixel size variations. We thus wanted to separate this long-term variation from any variations generated by the pixel size.

\begin{figure}[!htbp]
\includegraphics[width=\hsize]{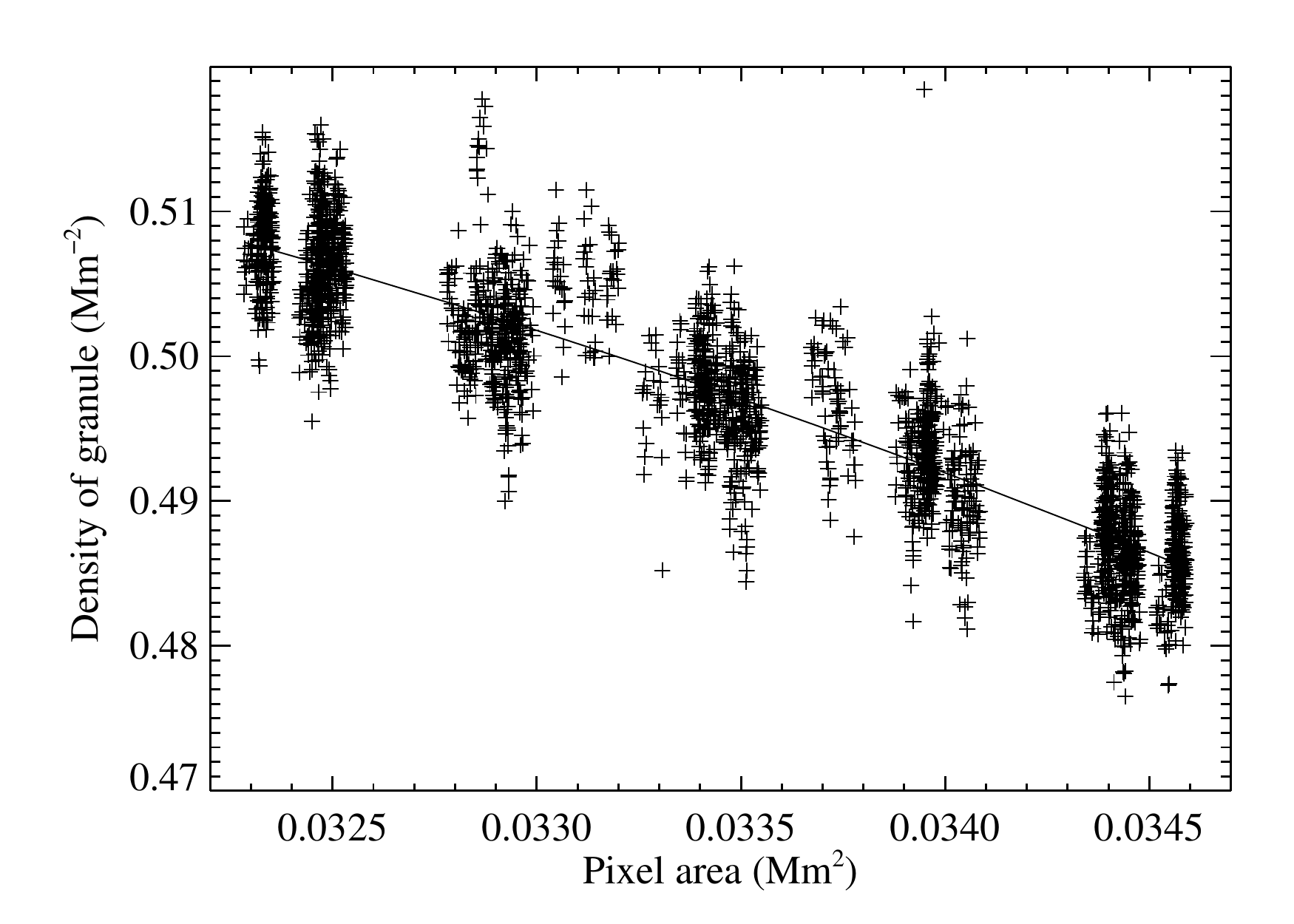}
\caption{Granule density as a function of the pixel area. The solid line is a fitted second-order polynomial.
\label{fig:density_pixel}}
\end{figure}

\begin{figure}[!htbp]
\includegraphics[width=\hsize]{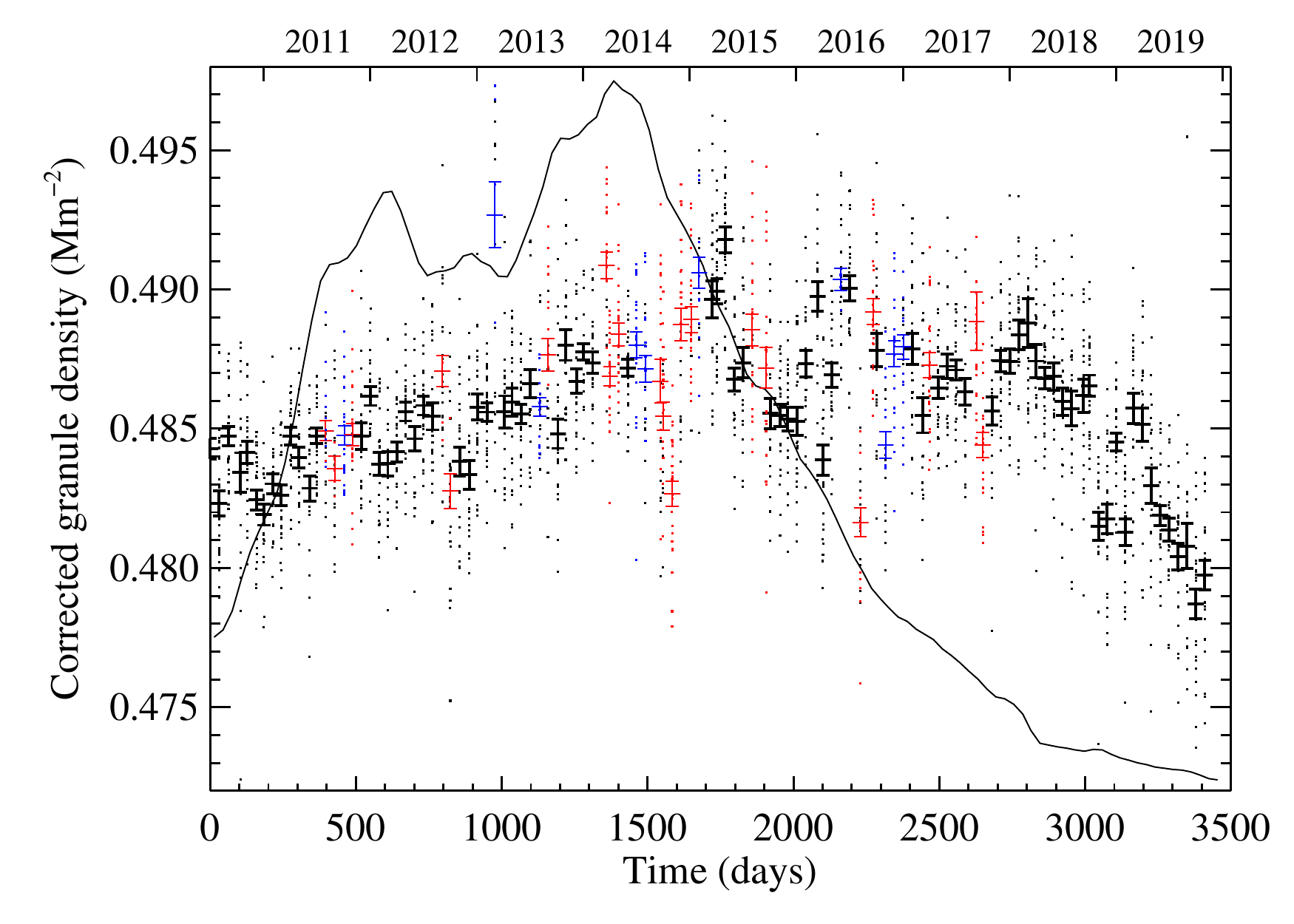}
\caption{Density of granules per Mm$^2$ corrected for the pixel area. Dots correspond to individual measurements, and crosses show daily averages with estimated error bars. We recall that we selected one single day for each month of the observation period. Dates where spots (pores) are present in the field of view are indicated with red (blue) symbols. Variations of the 13-month smoothed monthly mean total sunspot number \citep{sidc} are overplotted (solid line).
\label{fig:evol_densitycor}}
\end{figure}

 Figure~\ref{fig:density_pixel} shows the tight correlation between the pixel area and the measured granule density. We performed a second-order polynomial fit to model this smooth relation. We then used this fit to calibrate the granule density: We normalized the measured quantities with the fitted relation. Doing so, we needed to use an arbitrary reference value. As a consequence, only relative variations are meaningful. Figure~\ref{fig:evol_densitycor} displays the corrected granule density during the cycle.

For each selected day, we got hourly measurements. We thus have 24 values every day  (except for three days, for which one or two values are missing), which were averaged to compute daily mean values. We computed error bars of these means by assuming that daily standard deviations are representative of errors of individual measurements (typically $\sigma\approx 0.0022\:\mathrm{Mm}^{-2}$). To compute daily errors, we also assumed that each measurement is statistically uncorrelated to the others. This is ensured by the fact that two consecutive segmented images are significantly different since granule lifetimes ($8\sim 20'$) are shorter than the time between two measurements (1~hr).

We compared the evolution of granule density with the evolution of the 13-month smoothed monthly mean total sunspot number \citep{sidc}, which is used as a proxy for the solar magnetic activity (see the solid line in Fig.~\ref{fig:evol_densitycor}). We observe that the granule density increases with magnetic activity and that the granule density then decreases when the activity decreases. We notice that the maximum in density is delayed by $300\sim400$ days compared to the activity maximum. The beginning of the decay is slow, with a lot of fluctuations, but it drops markedly over the two last years and reaches its minimum values during the solar minimum. The amplitude of the variation is about 2\% (2.5\% peak-to-peak).

Finally, we wanted to ensure that these variations do not originate in the pollution of images with magnetic structures. Of course, when the activity is higher, there is a higher risk of finding spots, or simply pores, in our field of view. Such magnetic structures may pollute our estimates of the granule density. 
We thus went through all 2711 images to identify the dates where spots or pores appeared in our field of view. These dates are labelled with distinctive colours (red and blue) in Fig.~\ref{fig:evol_densitycor}. We thus confirm that the trends are unchanged when these points are discarded. We also notice that the part exhibiting the greatest dispersion corresponds to the period where the field is more often polluted by magnetic structures and that the strongest outliers actually correspond to images affected by magnetic structures.

\begin{figure}[!htbp]
\includegraphics[width=\hsize]{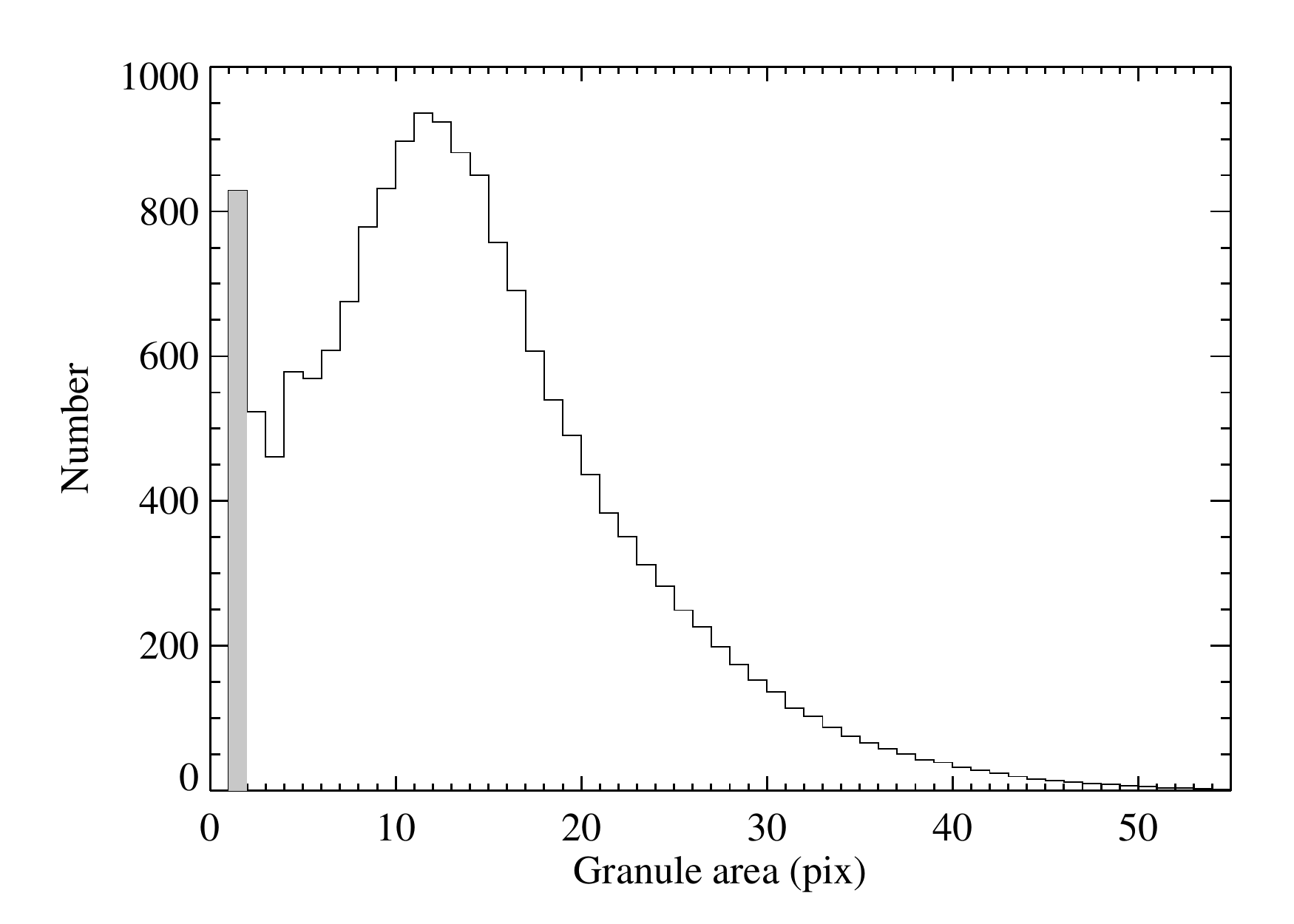}
\caption{Mean distribution of the granule area (in number of pixels) in the observed field of view on 1 July 2010. The shadowed bin corresponds to the one-pixel-size elements that are rejected. 
\label{fig:histo_area}}
\end{figure}

\subsection{Temporal evolution of the corrected mean granule area} \label{ssec:area}

From the segmented images, we also computed the mean area of granules. This is a complementary indicator that can be compared with the density, even if it is not fully independent. The density only relies on the count of segmented elements, and the mean area also takes their sizes into account. A representative histogram of granule sizes is plotted in Fig.~\ref{fig:histo_area}. As discussed in the previous section, the distribution is cut towards the small granules. As mentioned in Sect.\ \ref{sec:red}, we rejected one-pixel elements. We conducted complementary tests by also rejecting bigger granules (up to five pixels); the results of these tests did not call into question any of the conclusions presented in this section.

We thus computed the mean granule area in every image in pixels and converted it to Mm$^2$ using the real pixel size in megametres. As for computing the density, we carefully treated the boundaries of images: When we computed the mean granule area, we discarded the granules that are inside the boundary zone (see the definition in Sect.~\ref{ssec:evol_density}), and we considered granules only if at least half of their surface lay inside the image.
Once this was done, we performed exactly the same analyses as for the density.

\begin{figure}[!htbp]
\includegraphics[width=\hsize]{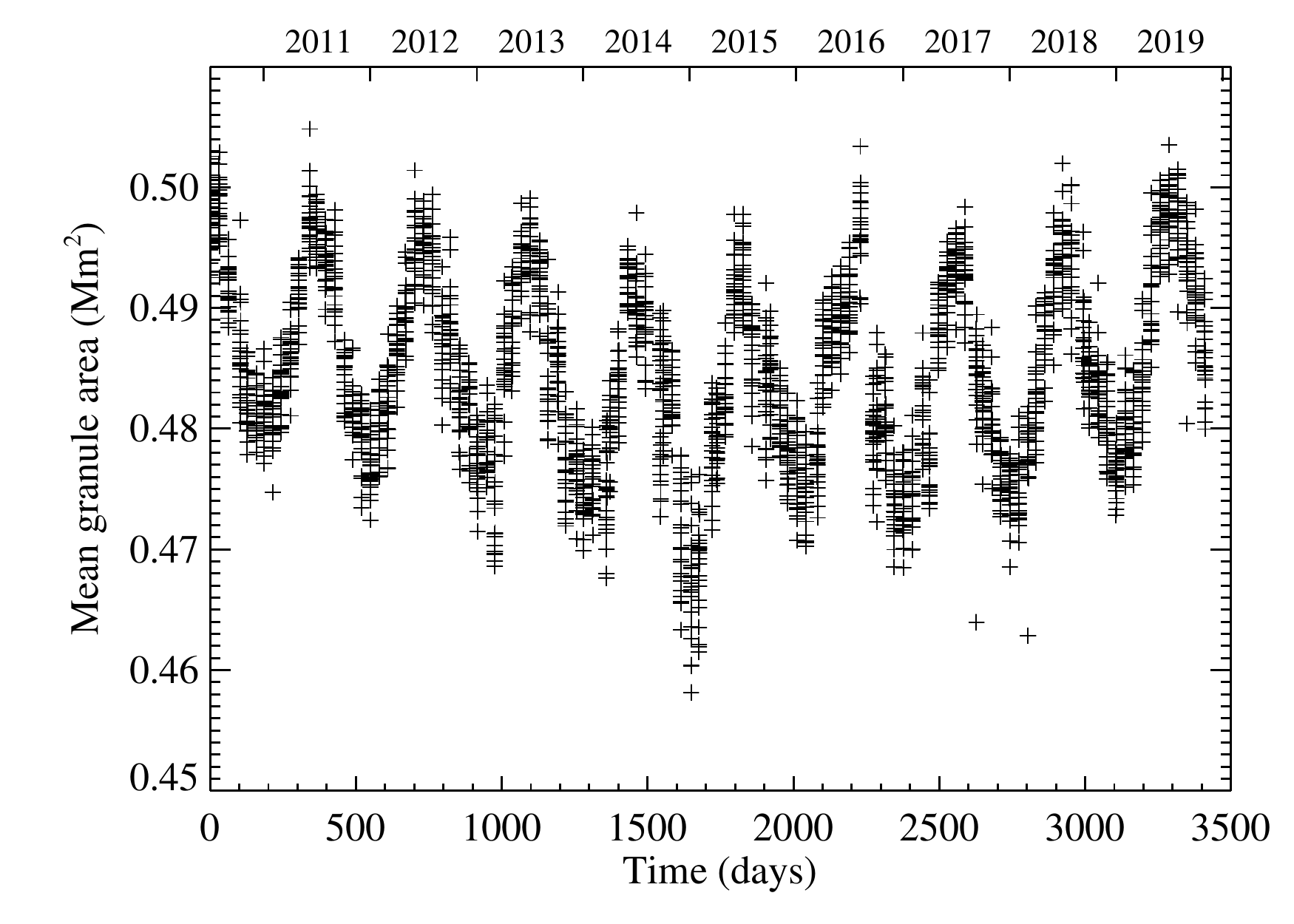}
\caption{Temporal evolution of the observed mean granule area measured at the disc centre with SDO. Each cross corresponds to the measurement obtained from an image. The time axis starts on 1 July 2010.
\label{fig:evol_area}}
\end{figure}

Figure~\ref{fig:evol_area} shows the temporal evolution of the mean granule area (similar to Fig.~\ref{fig:evol_density}). As for the density, the evolution is dominated by an annual oscillation -- due to variations in the pixel size -- superimposed on a global trend. All tests and the verification described in Sect.~\ref{ssec:evol_density} were conducted and led to the same conclusions.

\begin{figure}[!htbp]
\includegraphics[width=\hsize]{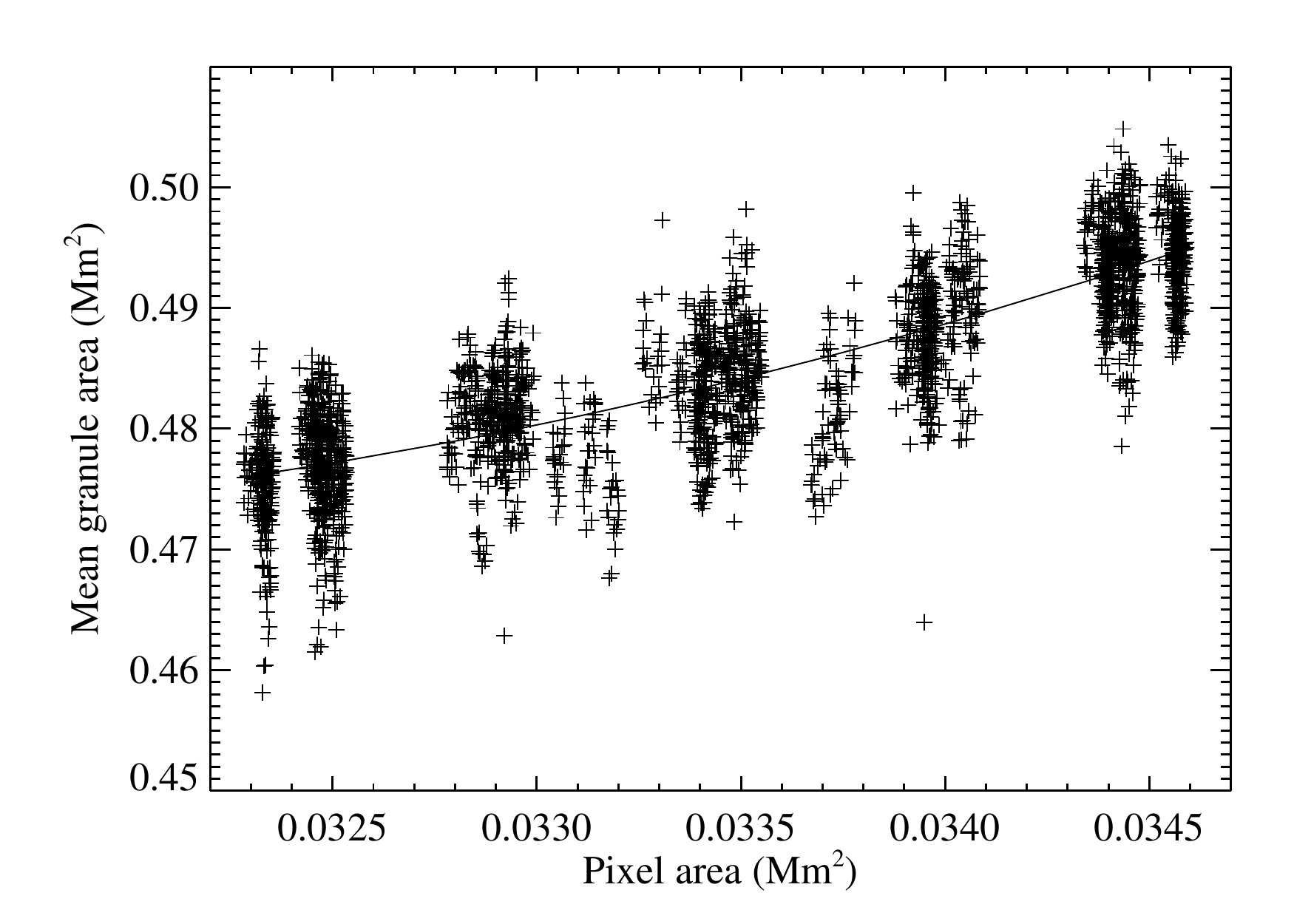}
\caption{Mean granule area as a function of the pixel area. The solid line is a fitted second-order polynomial.
\label{fig:area_pixel}}
\end{figure}

\begin{figure}[!htbp]
\includegraphics[width=\hsize]{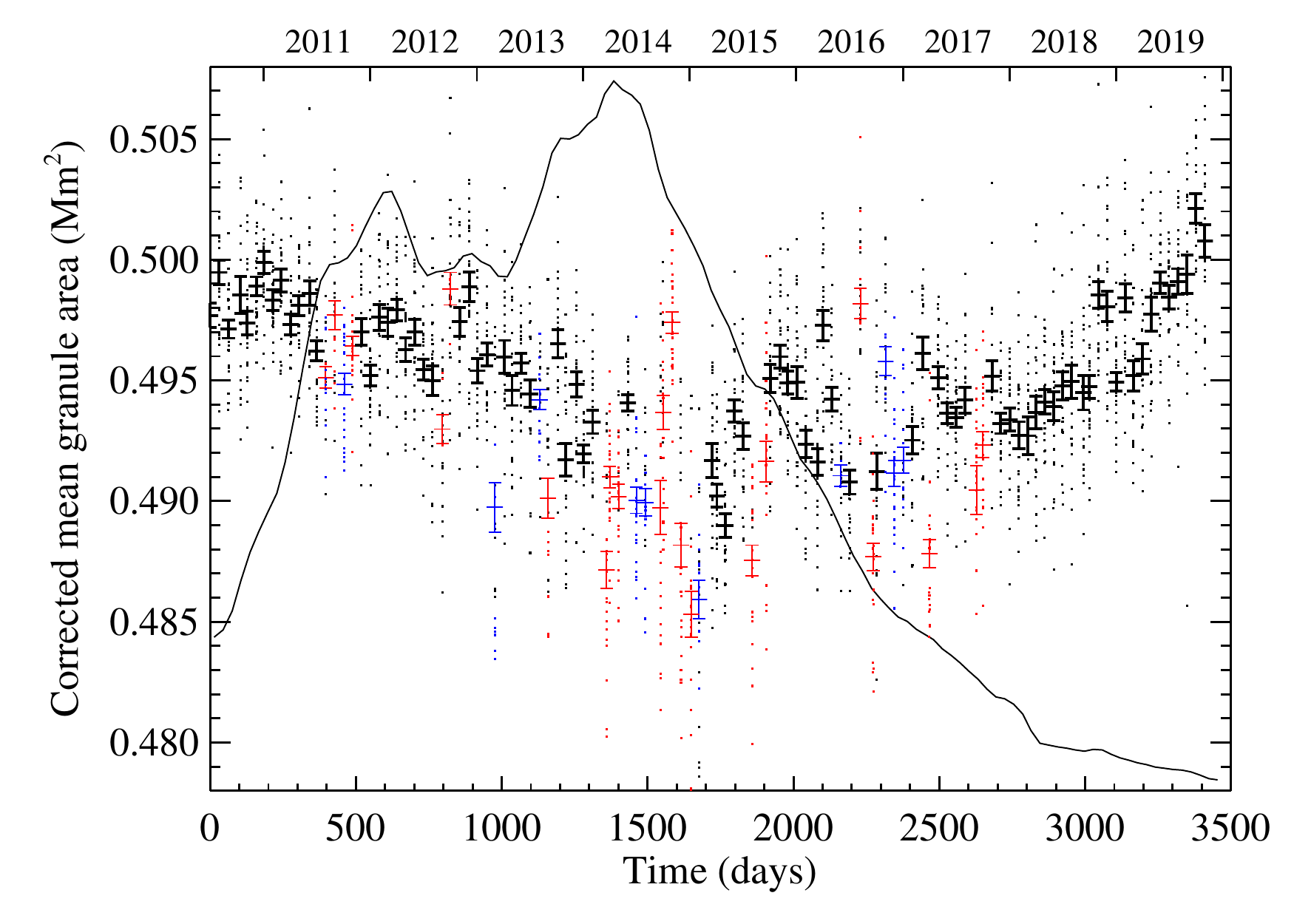}
\caption{Mean granule area corrected for the pixel area (same meanings as in Fig.~\ref{fig:evol_densitycor}).
\label{fig:evol_areacor}}
\end{figure}

Like the granule density, the measured mean granule area is strongly impacted by the pixel size. Therefore, we applied the same procedure to correct them. Figure~\ref{fig:area_pixel} shows the relation between the mean granule area and the pixel area (similar to Fig.~\ref{fig:density_pixel}). We fitted a polynomial to model this tight relation and used it to correct our measurements for this effect.

 Corrected mean granule areas are plotted in Fig.~\ref{fig:evol_areacor} (similar to Fig.~\ref{fig:evol_densitycor}). We also performed daily averages and compared their evolution to the sunspot number; the points that are potentially affected by spots and pores are identified with coloured symbols in the figure.
A very clear trend appears in this plot. This trend remains very clear even with measurements affected by magnetic structures discarded.
 The mean granule area decreases as activity increases and reaches a minimum on days $1700\sim1800$, which corresponds to the maximum in granule density; the mean granule area then increases again when the activity weakens. As for the density, in the second half the measurements start to increase with a lot of fluctuations, before increasing smoothly over the two last years.
 The variation amplitude is around  2\%  (about 2.5\% peak-to-peak if considering only a pure quiet Sun).

\begin{figure}[!htbp]
\includegraphics[width=\hsize]{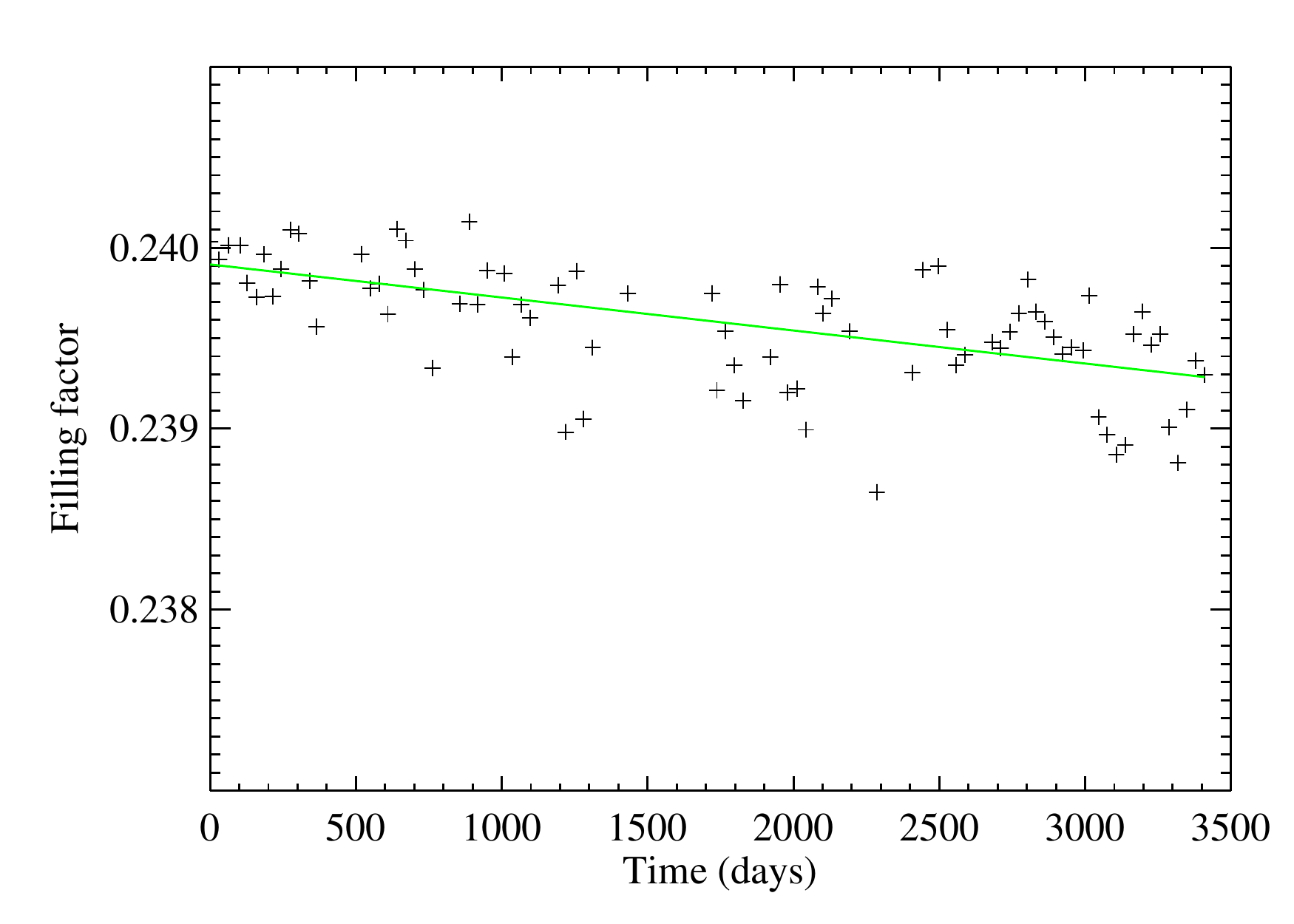}
\caption{Evolution of the measured daily-averaged filling factor of granules computed as the product of their mean area and density. Data points affected by spots or pores have been removed. A straight line indicates a linear fit.
\label{fig:evol_filf}}
\end{figure}
 We thus recover consistent results between the density and mean area of granules because we expect them to be anti-correlated. Indeed, we expect the filling factor of granules to be constant over time since the variations in irradiance due to granulation over the cycle is smaller than 0.01\% \citep[\eg][]{HUDWOOD1983}. The filling factor is nothing but the product of the density by the mean area and is plotted in Fig.~\ref{fig:evol_filf}. We recovered a flat profile of this product by excluding points affected by spots and pores. We nevertheless noticed a very slow and continuous decrease over nine years ($\sim0.2\%$) that is not correlated with the cycle. We attributed such a slow decay to instrument ageing (especially the ageing of detectors, electronics, optics, etc.), which may slowly degrade the image quality.

\section{Comparison with the \emph{Hinode} dataset}\label{sec:hinode}

To complement this analysis only relying on one instrument, we wanted to analyse high-spatial-resolution data from the SOT aboard the \textit{Hinode} spacecraft \citep[\eg][]{STISO08,ITSSO04}.
The SOT has a 50~cm primary mirror with a spatial resolution of about 0\farcs2 at a wavelength of 550\:nm. For our study,
we used blue continuum observations at 450.45\:nm from the \textit{Hinode}/SOT broadband filter imager (BFI).
We selected 191 observations over eight years (2008-2016) with a time step as regular as possible.
The selected sample is limited for mainly two reasons.
First, the dataset stops in 2016 due to the failure of the BFI camera. Second, a large part of the \textit{Hinode}/SOT observations were done within the G~band ($\sim$430\:nm), which cannot be exploited with our segmentation technique due to the presence of numerous inter-granular bright spots.

After flat field and dark corrections, SOT data were corrected for the constantly decreasing flux due to the degradation
of the optics over time. In order to limit this temporal variation, all image contrasts
were normalized while keeping the average of each image at their original values. The field of view was $2038\times1014$~pixels, giving 
$222\arcsec\times110.5\arcsec$ with a pixel size of 0\farcs1089. The transformation from arcsecond into kilometre takes the distance between the satellite and the Sun into account.

These observations overlap with our SDO series for about 5.6 years (from July 2010 to February 2016). Compared to SDO/HMI, \emph{Hinode} observations provide a greater spatial resolution: The optical resolution is five times better, and thus the images do not need to be deconvolved. On the other hand, the temporal sampling is ten times smaller and the exploited field of view twice smaller, drastically reducing the statistical sample.

\begin{figure}[!htbp]
\includegraphics[width=\hsize]{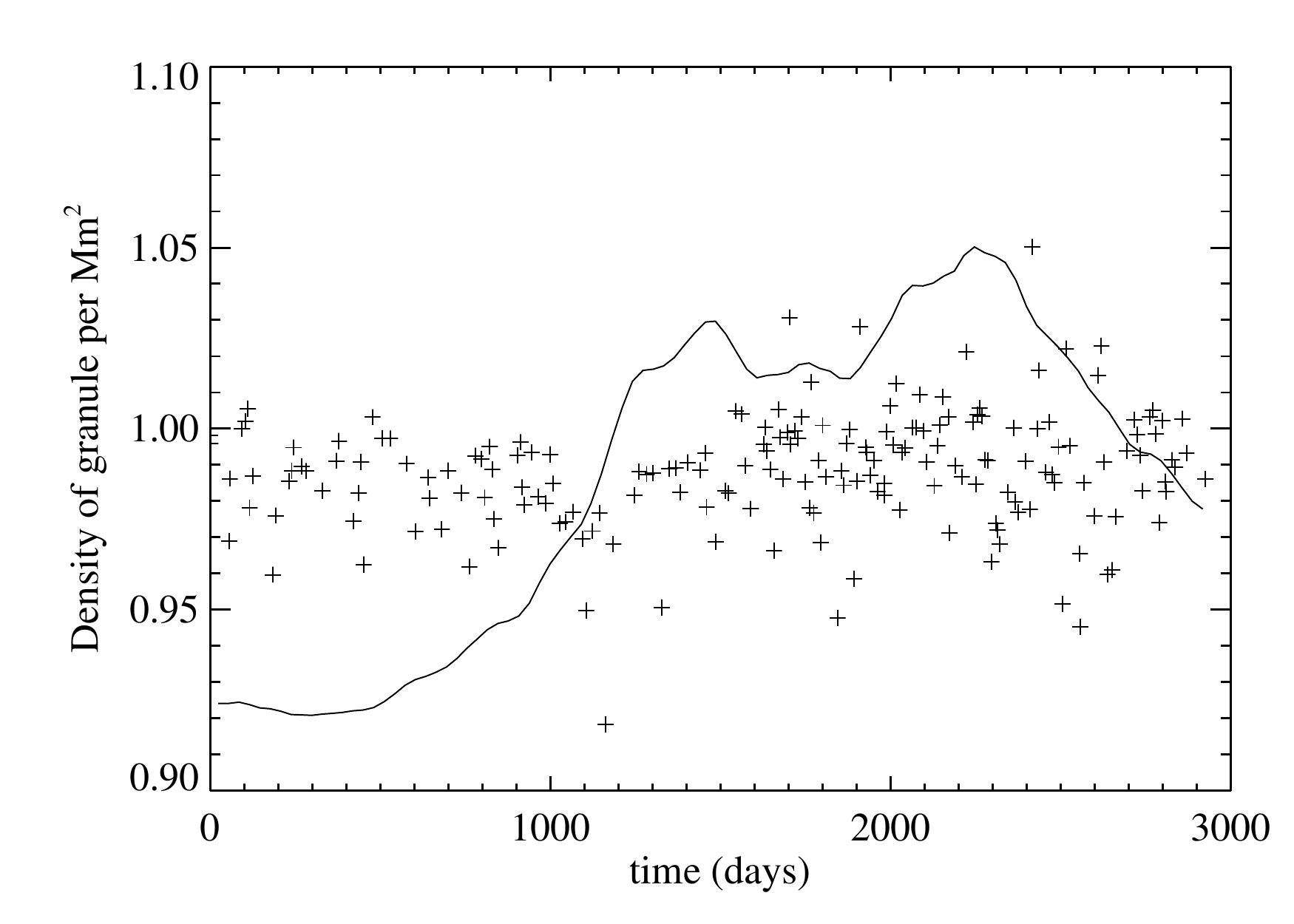}
  \caption{Temporal evolution of the granule density obtained from \emph{Hinode} observations. Density is corrected for the pixel area.  Variations in the monthly mean total sunspot number are overplotted (solid line). The time axis starts on 20 February 2008.
\label{fig:density_hinode}}
\end{figure}

\begin{figure}[!htbp] 
\includegraphics[width=\hsize]{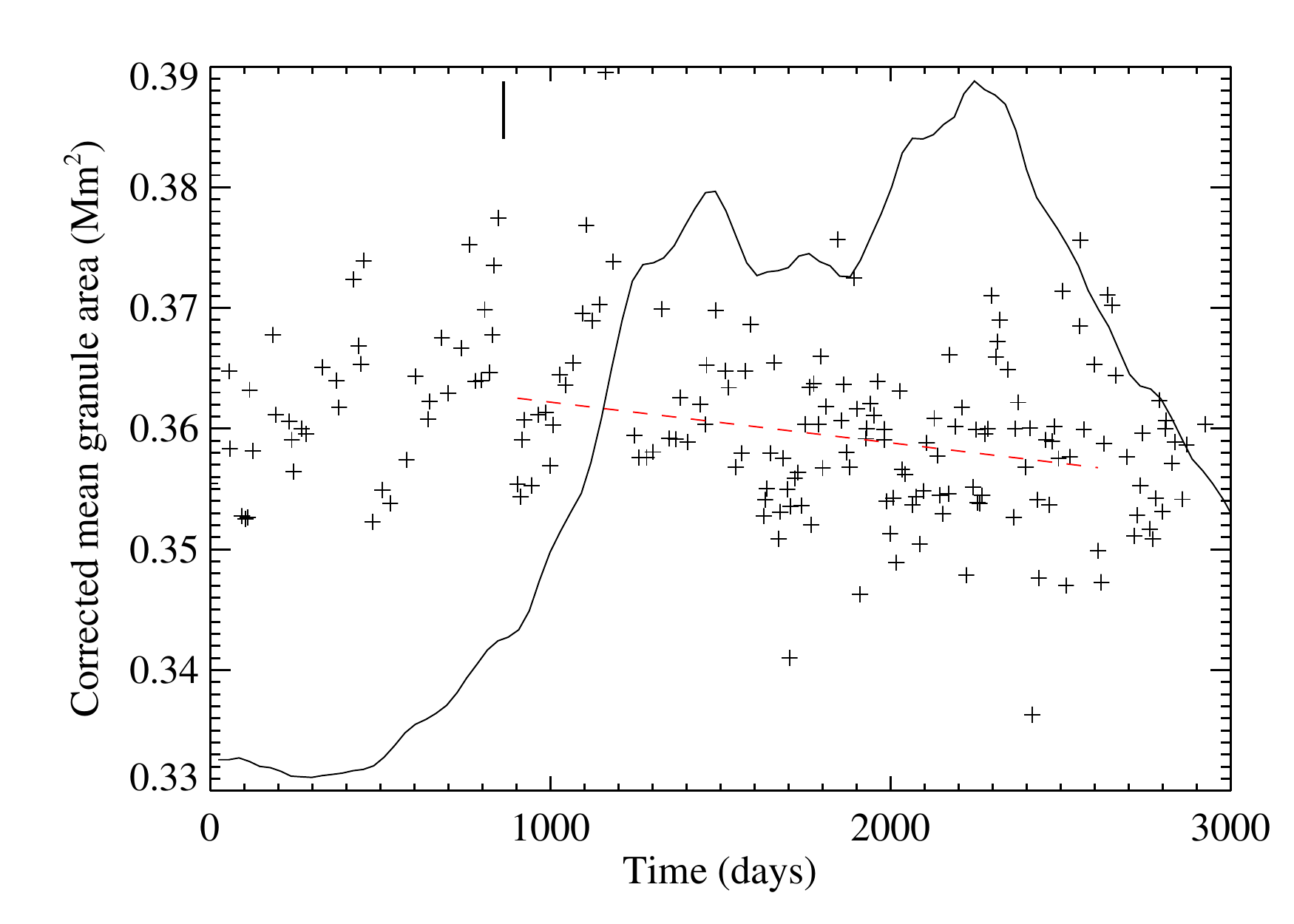}
\caption{Temporal evolution of the mean granule area obtained from \emph{Hinode} observations. The granule area is corrected for the pixel area. Variations in the sunspot number are overplotted (solid line). The straight dashed red line is a linear regression. The short, thick vertical line indicates the starting date of our SDO observation set (1 July 2010). The time axis starts on 20 February 2008.
\label{fig:area_hinode}}
\end{figure}

As done with the SDO data, the granule segmentation was performed by using convexity detection. We then applied exactly the same procedure as described above to compute the granule density and mean granule area. To be consistent, we also corrected them for pixel-size effects, despite the fact that the pixel size, due to the higher resolution, has a lot less influence. The corrected granule density and mean area are plotted in Figs.~\ref{fig:density_hinode} and \ref{fig:area_hinode}.

For both quantities, very large dispersion make it difficult to draw significant conclusions. However, while the granule density shows hardly any visible variation, a marginal trend appears in the mean granule area, which seems to decrease when activity increases.

The granule density seems noisier than the mean granule area. Indeed, with \emph{Hinode}
we observe many more small granules than with SDO thanks to its higher spatial resolution. The total number of granules detected with \emph{Hinode} (and hence the density) is more impacted by the large population of small granules than with SDO. However, the mean granule area appears to be less affected since we are averaging areas and not just counting them. We thus focused on this quantity and performed some statistical analyses.

As presented in Sect.~\ref{ssec:area}, the mean granule area decreases during the first $\sim$ 1750 days of the SDO/HMI time series. Hence, we analysed mean granule area observed with \emph{Hinode} on the same time interval (1750 days from 1 July 2010). We first needed to estimate the errors of our measurements. To do so, we performed a full analysis of a 24-hour sequence and computed the standard deviation. We thus estimated the relative error of measurements at 1.1\% (typically 0.004\:Mm$^2$).
Taking these errors into account, we performed a linear regression on that interval and recovered a decay (see Fig.~\ref{fig:area_hinode}). Interestingly, the decay is statistically more probable than a constant value: After comparing the $\chi^2$ of a linear model with that of a constant model, we rejected the latter with a $p$ value ${}\approx 10^{-5}$. From the linear fit, we derived that the mean granule area had decreased by $1.61 \pm 0.36\,\%$ over this time interval. We then performed the same regression over the same interval using SDO/HMI and found a decrease of $1.51 \pm 0.05\,\%$.
We conclude that both datasets are in agreement and that, despite being noisier, \emph{Hinode} observations confirm the change in the granulation scale detected with SDO/HMI on the considered interval.

\section{Impact of plages and the magnetic network}\label{sec:mag}

In previous sections we focus on the quiet Sun by excluding images polluted by spots or pores. Going a step further, we wanted see whether the observed variations are related to plage regions or to active network structures \citep[\eg][]{Foukal88} or whether granulation variations are still present in the inter-network. 

To achieve this objective, we took advantage of magnetograms that are simultaneously provided by HMI with each white-light image. From each magnetic map, we built a mask according to the following four-step procedure: (1) All pixels above a given threshold (30~G) were set in the initial mask. (2) We applied a closing operator (a dilatation followed by an erosion operation) to gather close structures. (3) We applied an opening operator (an erosion followed by a dilatation operation) to remove isolated pixels, which are generally generated by noise. (4) We finally applied a dilatation operation to provide a safety margin.
Steps 2 and 3 were introduced to avoid masks with a lot of unnecessary small holes or small islands.

\begin{figure}[!htbp]
\centering
\includegraphics[width=0.7\hsize]{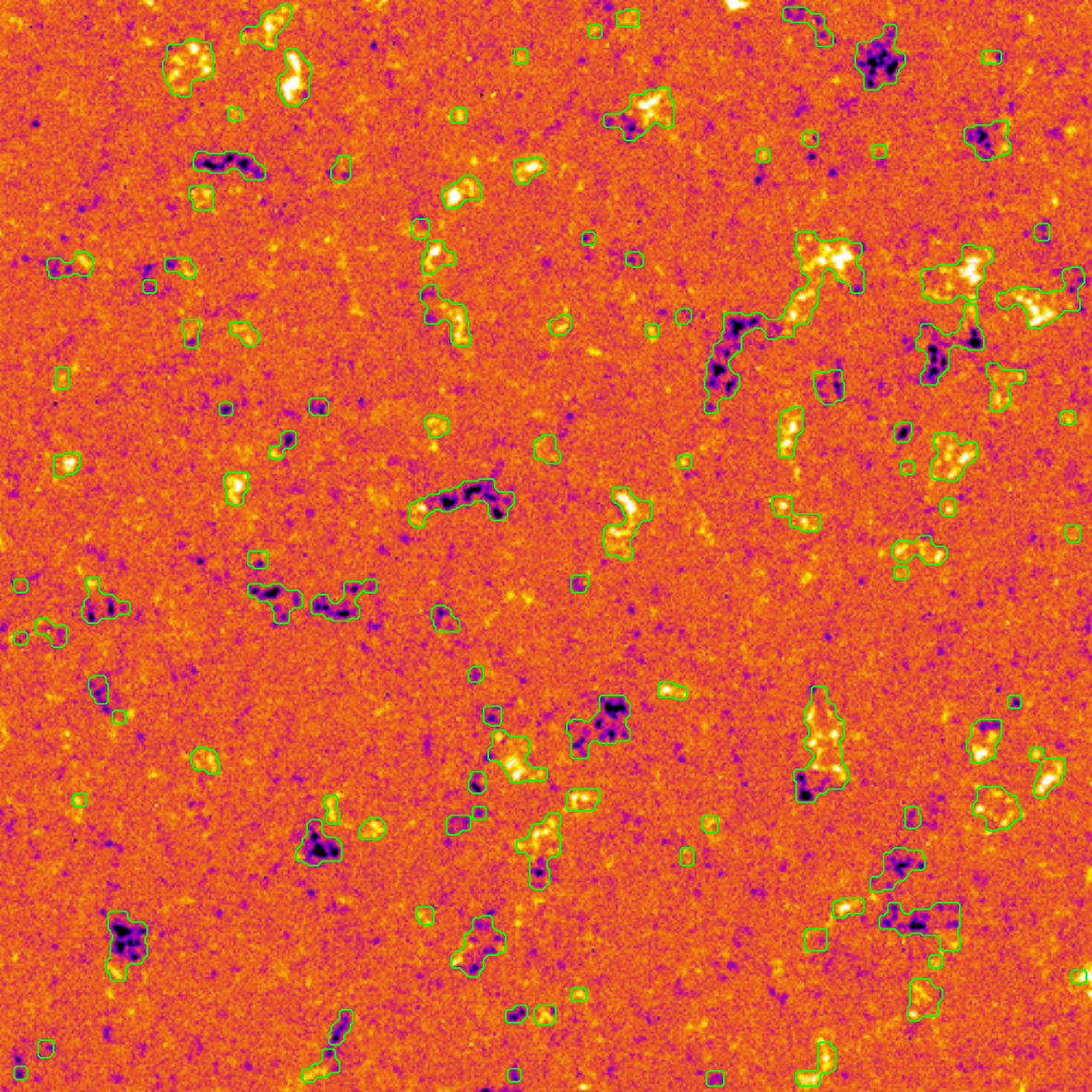}
\caption{HMI magnetogram corresponding to the first analysed field. The colour scale indicates the magnetic field strength from $-100$~G (purple and black) to $100$~G (yellow and white). Green contours show the borders of our mask.
\label{fig:mask}}
\end{figure}

Such a mask is shown Fig.~\ref{fig:mask}. We conducted various tests to build masks using slightly different methods (by skipping steps 2 and 3 or by slightly modifying the threshold from 20 to 50~G), but the conclusions of this section remained unchanged.

We then measured the density and mean granule area outside the mask using the same procedure as the one developed to properly deal with borders.
Nevertheless, we wanted to ensure that our masks did not introduce spurious variations. Masks with a lot of structures may affect the measurements more than masks with fewer structures. We used some artificial observations (derived from simulations; see Sect.~\ref{ssec:pix}) to which we applied all our masks. In this artificial observation, as the granulation distribution is statistically spatially homogeneous, we had to recover a consistent measured density and mean granule area with all of the masks.
We actually do not find any significant bias in the measurements of the granulation density or the mean area made with any of the masks.

\begin{figure}[!htbp]
\centering
\includegraphics[width=\hsize]{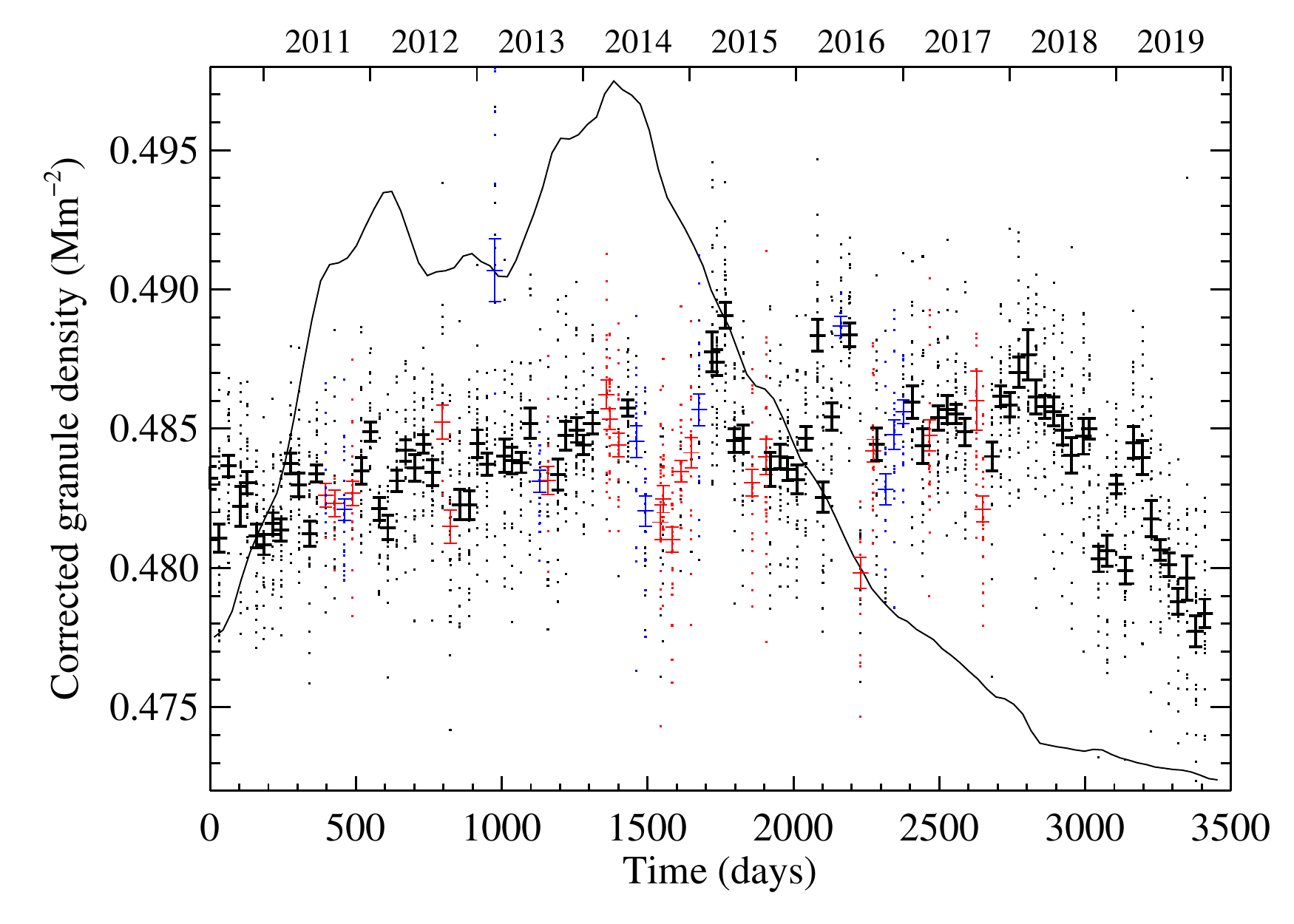}
\caption{Density of granules per Mm$^2$ corrected for the pixel area in the inter-network. This figure is similar to Fig.~\ref{fig:evol_densitycor}, but masks have been applied to remove regions with a magnetic field stronger than 30~G. Fig.~\ref{fig:mask} shows such a mask.
\label{fig:evol_densitycor_intra}}
\end{figure}

\begin{figure}[!htbp]
\centering
\includegraphics[width=\hsize]{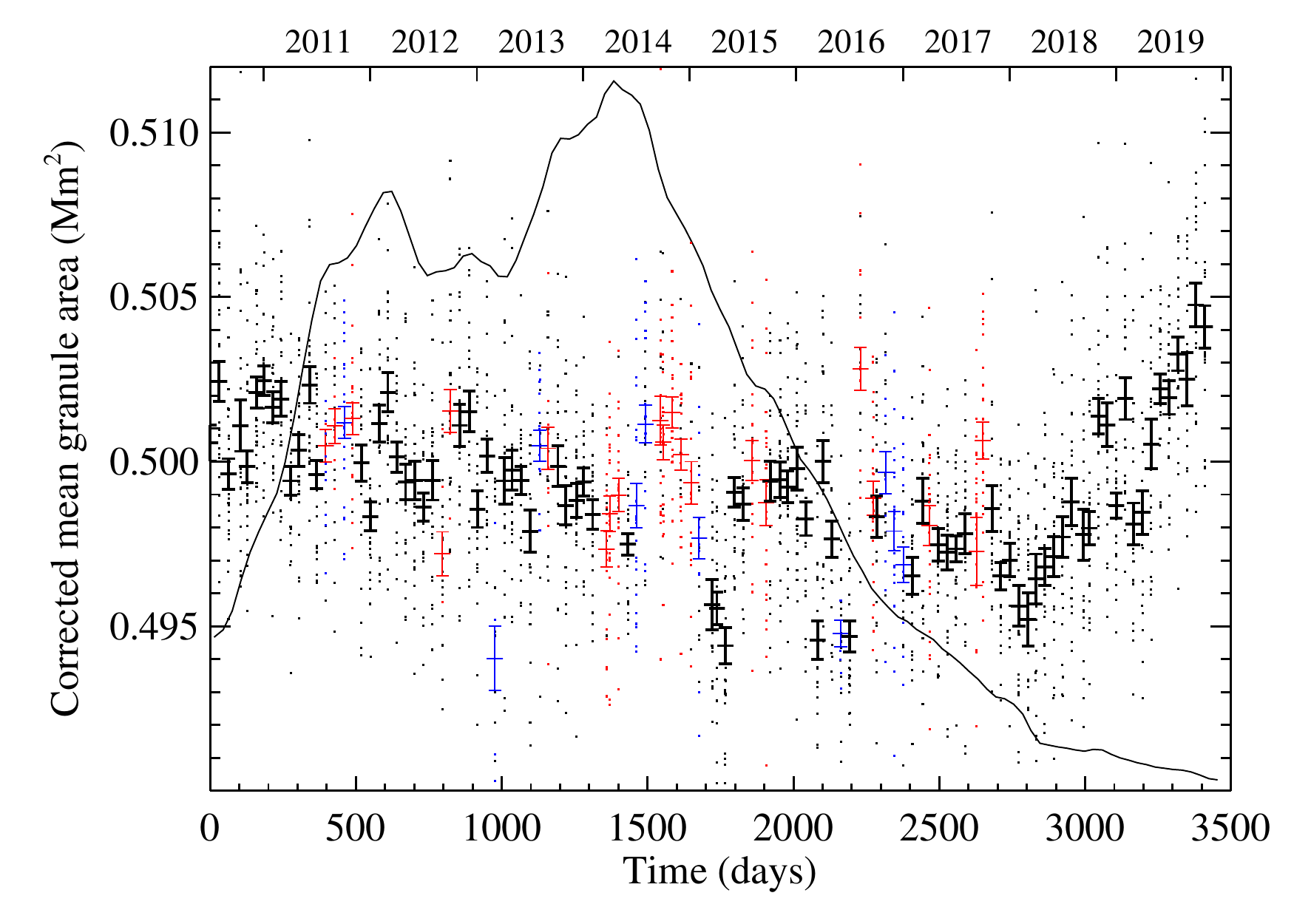}
\caption{Mean granule area corrected for the pixel area in the inter-network. This figure is similar to Fig.~\ref{fig:evol_areacor}, but masks have been applied to remove regions with a magnetic field stronger than 30~G. Fig. \ref{fig:mask} shows such a mask.
\label{fig:evol_areacor_intra}}
\end{figure}

We applied masks to our real HMI dataset, measured the granule density and mean area, and corrected them for pixel-size effects. Results are shown Figs.~\ref{fig:evol_densitycor_intra} and \ref{fig:evol_areacor_intra}, which can be compared to Figs.~\ref{fig:evol_densitycor} and \ref{fig:evol_areacor}.
First, the density is in average slightly lower than in the full frame. This is consistent with theoretical expectations: Where the magnetic field is weaker, granules are expected to be larger. We also notice that the measured mean area appear slightly noisier than the density. Second, the variations observed along the solar cycle are weaker but still present. We can conclude that although small magnetic structures of the network explain a part of the variations reported in previous sections, there are also variations in the inter-network regions.

We can also use dedicated masks to clean spots and pores as well as plages from the images. We made this effort and present the results in Appendix~\ref{app}. They are fully consistent with what is presented in Sects.~\ref{ssec:density} and \ref{ssec:area}.

\section{Conclusion}\label{sec:conc}

In this paper we report the detection of variations in granule density and mean area over the period 2010 - 2019. We conclude that, once annual spurious effects are corrected for, a variation of about 2\% in the granulation scale occurs during the solar cycle,
the granules being the smallest about one year after the Sun reaches its maximal activity.
These changes are detected clearly with HMI/SDO, and \emph{Hinode} observations show consistent and significant, if noisier, variations. We also showed that while small magnetic structures of the network may explain a part of these variations, we still observe variations in the inter-network regions.

We stress that we only recovered relative variations; the absolute values we report are not meaningful for several reasons. First, the granule detection technique we have used is very robust but underestimates the granule surface. Second, we miss the smallest granules due to our limited resolution. As a side effect, our absolute measurement is sensitive to the pixel size (which changes during the SDO orbit) because the cutoff of the observed granule size distribution is consequently moving. Luckily, we were able to correct for the impact of the pixel size and recover long-term variations. Nevertheless, we must keep in mind that, because of this ad hoc correction, variations with periods shorter than one year (\ie\ the orbital period) must be treated cautiously and may still be polluted.

We carefully verified whether the instrument or the analysis method could have generated the observed variations. Of course, despite our careful inspection, we may have missed a hidden parameter; furthermore, we have not addressed the possible effects of ageing, which leads to a lower mean contrast of HMI images. Even though we verified that a change in contrast alone does not affect our segmentation method, this could be a sign of an underlying degradation in image quality or sharpness over time. Such a degradation would affect our segmentation method as de-focussing would (see Sect.~\ref{ssec:evol_density}).
The slow decay observed in the filling factor is probably a sign that a degradation has been occurring. Here is one possible explanation: With time, as images become (slightly) more degraded, in similar situations we would less easily detect the small granules, which may reduce the filling factor. Of course, this should also affect the measured density -- which should decrease -- and mean area -- which should increase. Nevertheless, we are confident that the ageing does not provoke the reported 2\% variation since the impact of ageing is monotonic with time, whereas the observed granulation scales decrease then increase again. Thus, even if we would have to correct the evolution of mean area (respectively, density) for a global increase (respectively, decrease) due to ageing, the V-shape of the curve would remain and only the amplitude would be (slightly) affected.

We report, for the first time, a direct measurement of granule variations in the quiet Sun along the cycle; indirect clues of global modifications in granulation (integrated over the whole Sun) had already been suggested by variations in Fraunhofer line bisectors \citep[\eg ][]{Livingston82,Livingston99}. Moreover, previous works indicated that variations in granulation during the cycle could be only of weak amplitude \citep{Muller2018}; here we were able to sufficiently lower our detection level thanks in particular to the high stability and quality of the SDO/HMI instrument.

Effects of magnetic fields on granulation have already been observed on the Sun: for example, it has been established that granules in magnetic plages are smaller than those in the quiet Sun \citep[\eg][]{Title92,Narayan10}. Indeed, it can be shown that convective cells are shrunk in a vertical magnetic field because the critical horizontal wavelength of convective instability lowers when a vertical magnetic field grows \citep[\eg ][]{chandra61}. Thus, a global change in the solar poloidal magnetic field may affect the granule size. Nevertheless, the interactions between solar convection and the magnetic field are complex and subtle and may occur at the global scale or locally at the granule scale. Quantifying the changes in terms of magnetic fields would require detailed modelling that is beyond the scope of this study.
This paper brings two new observational constraints for magnetohydrodynamics  models: the amplitude of the change and the $\sim$1-year phase shift relative to classical activity indicators, such as the sunspot number.

In this paper we have described the evolution of the granule properties at the disc centre (\ie\ at low latitude only). It would be very useful to extend this study to higher latitudes to verify whether the trend continues. Indeed, the latitudinal structuring of the solar magnetic field evolves significantly over a cycle. Therefore, mapping the granulation scales both in time and in latitude is perfectly relevant. Due to projection effects, segmentation processes at high latitude may be more delicate and require more care. This will be the subject of a future study.

For stars other than the Sun, it is possible to recover global information about granulation through temporal photometric observations by computing indexes such as the `8 hr Flicker' \citep{Bastien13} or the so-called FliPer \citep{Bugnet18}, or simply by extracting granulation timescales and amplitudes from power spectra \citep[e.g.][and references therein]{GarciaBallotLR}. \citet{Santos18} studied the temporal variations in granulation timescales in \emph{Kepler} solar-like stars. They did not find a general link between activity and granulation for most of the stars.  For one star (KIC 10644253), characteristic granulation timescales show variations correlated with magnetic activity indexes. However, they report variations in granulation timescales of around 20\% that are directly correlated to the activity. This is the opposite of our present results, which also have a much lower amplitude. Nevertheless, we must remember that these \textit{Kepler} observations are averaged over the whole stellar photosphere and are not only of quiet areas.
Beyond solar physics, our objective is to better understand stellar surface convection and its interactions with magnetic fields, using the Sun as a close plasma physics laboratory.

  \begin{acknowledgements}

    This work was granted access to the HPC resources of CALMIP under the allocation
2011-[P1115]. We are grateful to SDO/HMI and \emph{Hinode} teams. Sunspot data are from the World Data Center SILSO, Royal Observatory of Belgium, Brussels (\url{http://www.sidc.be/silso/}).
We thank the anonymous referee for her/his insightful comments, which have improved the paper.
\end{acknowledgements}

  \bibliographystyle{aa}    
\bibliography{biblio}

\appendix

\section{Masking spots, pores, and plages}\label{app}

The masks we developed in Sect.~\ref{sec:mag} in order to isolate the magnetic structures of the solar network can be applied to remove spots and pores, as well as plages, in images that are polluted by them.

\begin{figure}[!htbp]
\includegraphics[width=\hsize]{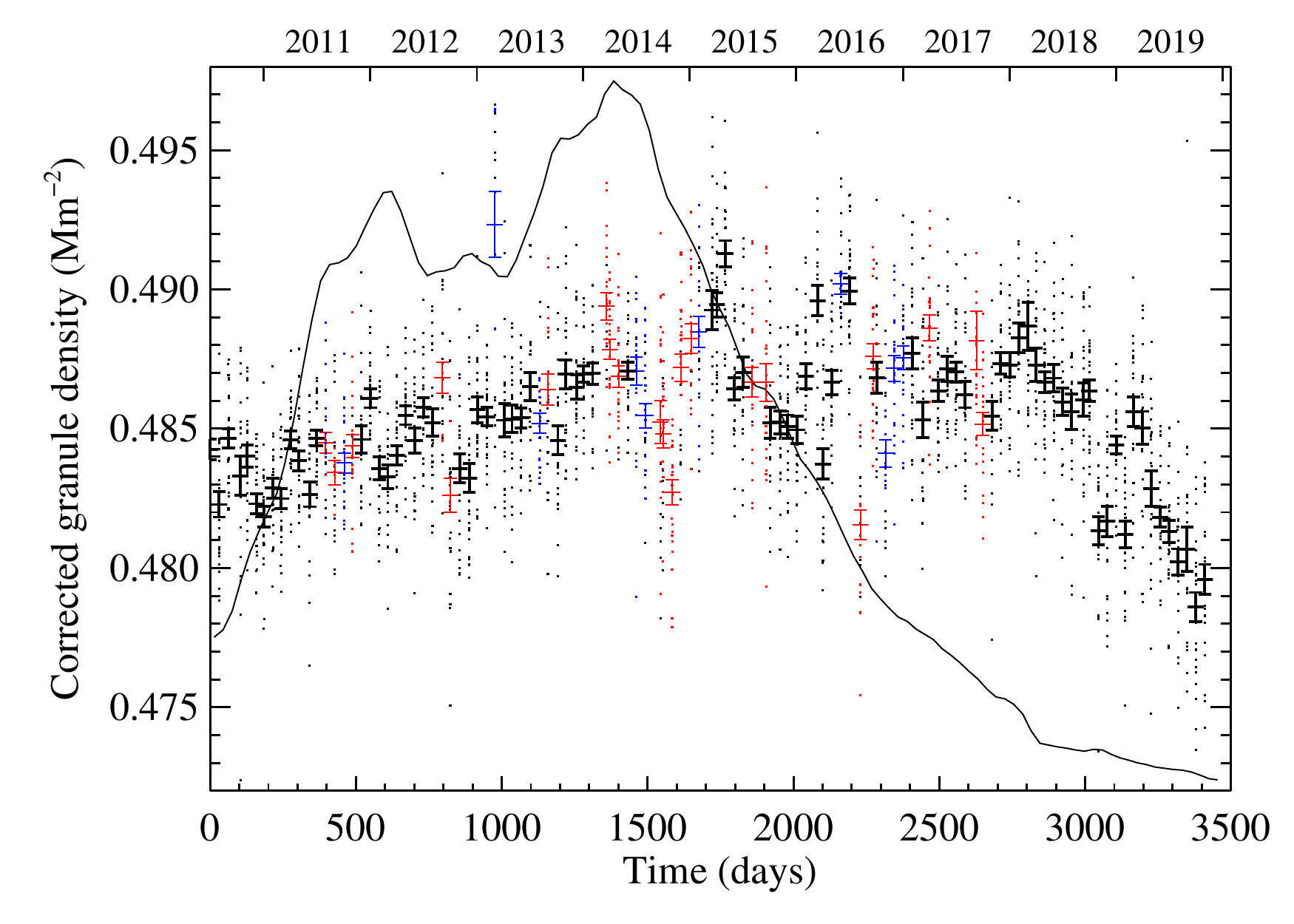}
\caption{Density of granules per Mm$^2$ corrected for the pixel area. This figure is similar to Fig.~\ref{fig:evol_densitycor}, but masks have been applied to remove structures with a magnetic field stronger than 200~G.
\label{fig:evol_densitycorM200}}
\end{figure}

\begin{figure}[!htbp]
\includegraphics[width=\hsize]{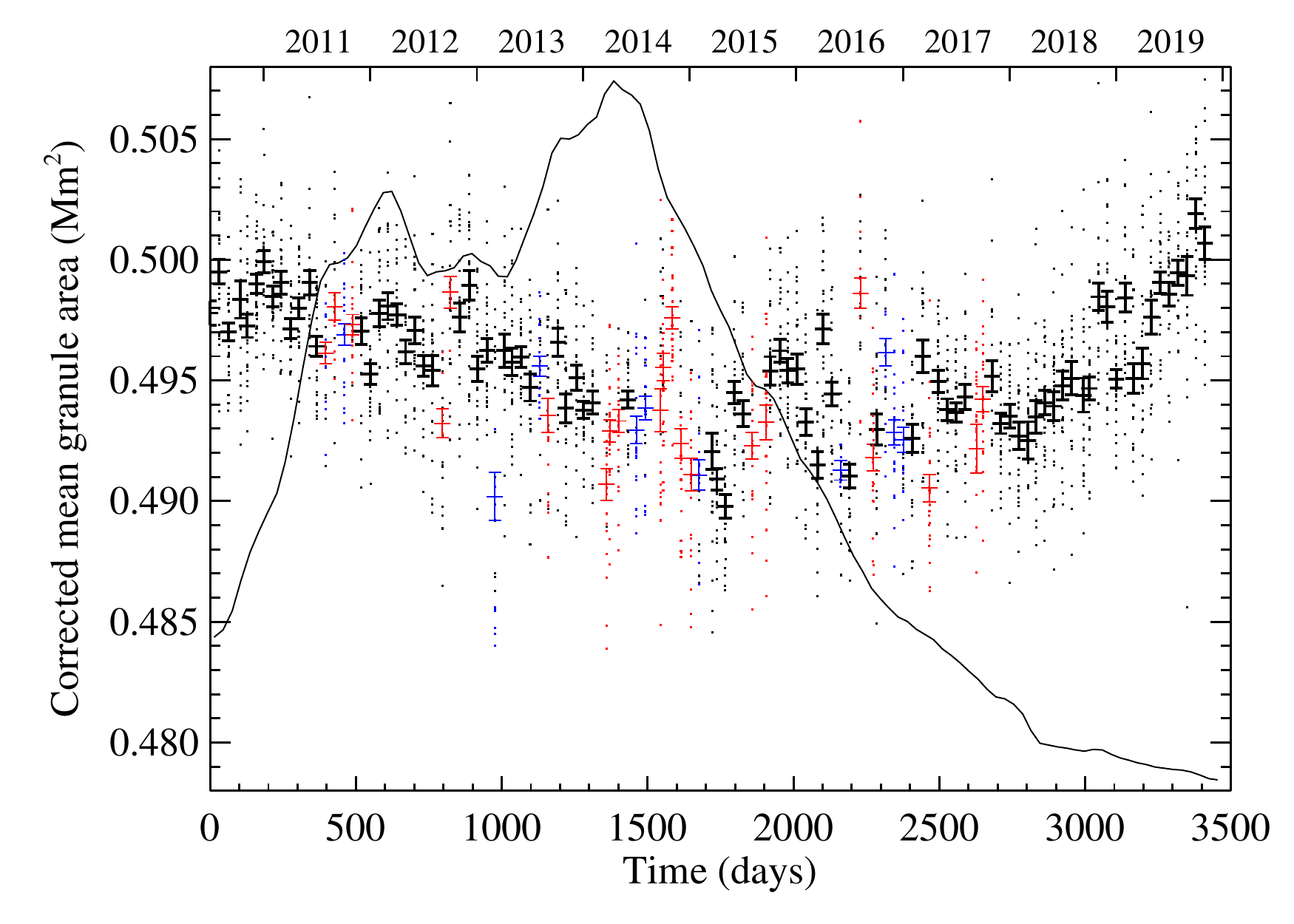}
\caption{Mean granule area corrected for the pixel area. This figure is similar to Fig.~\ref{fig:evol_area}, but masks have been applied to remove structures with a magnetic field stronger than 200~G.
\label{fig:evol_areacorM200}}
\end{figure}

We thus constructed masks to remove magnetic structures above 200~G \citep[this threshold was chosen according to literature, e.g.][]{Kahil19} and applied our full procedure to measure the density of granules and the mean granule area. Results are shown in Figs.~\ref{fig:evol_densitycorM200}
and \ref{fig:evol_areacorM200}, which can be directly compared to Figs.~\ref{fig:evol_densitycor} and \ref{fig:evol_areacor}, respectively. As expected, the black symbols, which correspond to images that are tagged as containing no spots or pores, are practically the same, whereas blue and red symbols are less scattered in  Figs.~\ref{fig:evol_densitycorM200}
and \ref{fig:evol_areacorM200} compared to Figs.~\ref{fig:evol_densitycor} and \ref{fig:evol_areacor}.
The improvement is mainly visible for the evolution of the mean area.

\end{document}